\documentclass[11pt]{article}
\usepackage{latexsym}
\usepackage{amssymb}
\usepackage{amsmath}
\def\longbar#1{#1\kern-0.7em\raise1.3ex\hbox{{$-$}}}
   \def\d{\delta}
   
\def\m{\mu} \def\n{\nu}

\def\e{\varepsilon}  
\def\pa{\partial}

\def\be{\begin{equation}}
\def\ee{\end{equation}}
\def\bea{\begin{eqnarray}}
\def\eea{\end{eqnarray}}
\def\pA{\ensuremath{\partial\!\cdot\!A}}
\def\mes{\ensuremath{\int\!\!}}

\def\inv{{^{-1}}}

\def\tr{\ensuremath{\mathrm{tr}}}

   \def\cD{{\cal D}}

  \def\cO{{\cal O}}

\def\Amin{\ensuremath{\parallel\!\!A\!\!\parallel^2_{{\mathit min}}\,}}
\usepackage{graphicx}
\usepackage{dcolumn}
\usepackage{bm}

\begin{document}
 \title {The  Non-Local Massive  Yang-Mills Action  \\
as a 
\\Gauged Sigma Model
}

\author{
Mboyo Esole
\footnote{E. \ Mail: esole@aims.ac.za, esole@lorentz.leidenuniv.nl}\\
{\small  African Institute for  Mathematical Sciences,}\\
{\small 6 Melrose Road, Muizenberg, 7945  Cape Town,   RSA}\\
{\small and }\\
{\small Instituut-Lorentz, Universiteit Leiden,}\\
{\small  P.~O.~Box  9506, 2300 RA Leiden,  The Netherlands}
}
\maketitle

\begin{abstract}
We  show that the  massive Yang--Mills action  having as a  mass term the non-local operator  introduced by Gubarov, Stodolsky, and Zakharov is classically equivalent to a  principal gauged  sigma model.
  The non-local mass corresponds   to the topological term of the sigma model. 
 The latter is obtained once the degrees of freedom  implicitly generated in the   non-local action are explicitly implemented as group elements. The non-local action is recovered by integrating out  these group elements. 
 In contrast to the usual gauge-fixed treatment, the sigma model point of view  provides a safe framework in which calculation are tractable while keeping a full control of gauge-invariance. 
 It shows that the non-local massive Yang--Mills action is naturally associated with the low-energy description of QCD in the  
 Chiral Perturbation Theory   approach. Moreover, the sigma model   admits solutions called center vortices familiar in  different (de)-confinement and chiral symmetry breaking scenarios. This suggests that the 
 non-local operator  introduced by Gubarov, Stodolsky, and Zakharov  might be sensitive to center vortices configurations.
\end{abstract}
\newpage 
\section{Introduction}
Non-locality appears in a wide range of areas from  astrophysics and   cosmology to  string theories  and   non-commutative theories \cite{Pais:1950za,Eliezer,Simon:ic,Gomis:2000gy,Moeller:2002vx,Soussa:2003re,Seiberg:1999vs}. However,  it can  frighten the basic principles of a relativistic theory. The   assumption of  locality pervades most methods of 
 quantum field theory and  is  related to   unitarity, causality  and  renormalizability \cite{Bogolyubov,Barnich:th,Barnich:2000zw,Buchholz:1999xm}. It is then  far from obvious that a given  non-local theory  can make sense physically.

In this letter we are interested in the non-local mass generation scheme  developed  recently for Yang--Mills theories in the Landau gauge \cite{Kondo:2001nq}. 
The interest in non-local mass generation scenarios is mainly due to the absence of  local gauge-invariant  mass term in pure Yang--Mills theories in four dimensions \footnote{ This is a general feature of Yang--Mills theories in any number of dimensions.  However, in three space-time dimensions  the Chern--Simons term provides a topological term. \\
 We recall that \cite{Barnich:2000zw} a local function is a polynomial in the fields and a finite number of their derivatives all evaluated at the same space-time point. A local functional is the integral of a local function. This definition of locality is valid for theory involving only point-wise objects and is very practical for calculations. It is  based on  properties of distributions having support on a unique space-time point \cite{Bogolyubov}. From this perspective, theories involving extended objects are classified as   non-local. But a   more appropriate definition of locality  exists to deal with extended objects \cite{Buchholz:1999xm}.}
.   
Consequently,   a gauge-invariant massive Yang-Mills
 action requires an 
  extension of the number of fields,  like in the Higgs mechanism,  or 
   the introduction of non-local quantities.  
The Higgs mechanism is not appropriate when there is no room for new particles in the spectrum of the theory. Therefore, although non-locality is a tricky concept in quantum field theory, it naturally appears in mass generation without Higgs fields.  

A  non-local massive Yang--Mills model  has emerged recently \footnote{ See \cite{Kondo:2001nq} and references therein.} following the introduction by  Gubarov, Stodolsky, and Zakharov \cite{Gubarev:2000eu} of a new gauge-invariant operator having the dimension of a mass term. That operator is defined as the minimum  along the gauge orbit of the square potential $\int A^2$ and is usually denoted $\Amin$ (`$A^2$ minimum'). 
 Gauge-invariance  is  ensured  in principle since  the  minimum is the same for any two points on the same gauge orbit. But the price to pay is the loss of locality.   A  non-local massive action is obtained by adding \Amin to the Yang--Mills action. 
 \Amin is  a Morse functional  which contains  topological information on the gauge-orbit \cite{Fuchs:1995mb}. These type of functionals  obtained  from minimizing a given local quantities along the gauge orbit reveals interesting  properties of the structure of the configuration space of Yang--Mills theories. 
 The analysis of \Amin for example, requires a careful look at issues related to    Gribov copies \cite{Fuchs:1995mb,Stodolsky:2002st}. 
It also   emphasises  the role of the     the  center  of the Lie group to understand the structure of the configuration space, the distinction between   reducible and irreducible  gauge potentials and the relevance of the stratification of the configuration space   \cite{Fuchs:1995mb}.  It is not very surprising that a mass term is closely related to the topology. For example in three dimensions, the Chern-Simons term provides important topological information. \Amin is supposed to give as well information on the topological structures of gauge theories. This has been illustrated by lattice simulations which seems to indicate that   \Amin is sensitive to phase transitions in compact QED \cite{Gubarev:2000eu}. It was also argued on the lattice that the expectation value of the square potential is associated to instantons in the Landau gauge  \cite{Boucaud:2002nc}. All these lattice results are not understood in the continuous formulation of Yang--Mills theory.

It is convenient to think of the non-local massive   Yang--Mills action  as a (low-energy) effective action. The 
 possible  dangerous effects  associated with  non-locality 
\cite{Pais:1950za,Eliezer,Simon:ic,Gomis:2000gy,Moeller:2002vx}  are then not necessarily  the  result of fundamental violations of first principles. They  can  follow from  the  approximative  nature of the model \cite{Simon:ic}. 
In particular, unitarity and causality  might be preserved under a
certain  cut-off  and   renormalizability is not an issue. 
Since Yang--Mills theories have a  mass gap, an  effective massive  theory   provides an accurate  description of the physics at low energy. 
Another point of view is to treat the massive action as an effective action used for example to renormalize  the operator \Amin.

The functional \Amin is difficult to analyze  in a gauge-invariant way because of  its  unconventional definition. 
Therefore, the associated non-local massive  action is often defined by the trivial massive  action $\int F^2+m^2 A^2$  considered as a  gauge-fixed action  in the Landau gauge. Indeed, the square potential ($\int A^2$) reaches its local  minima  when it is
 transverse and this corresponds to the Landau gauge condition  ($\pa \cdot
 A=0$) \cite{Stodolsky:2002st}. 
At first sight, the gauge-fixed action in the Landau gauge seems to be  easier to handle  because of its locality. However, this type of `locality' is   a pure gauge-fixed mirage that can lead  to erroneous conclusions  when the usual techniques of (local) quantum field theory are used without checking their relevance in a non-local context 
\footnote{For example, the observables of the theory are usually identified with the BRST cohomology. But the gauge-fixed BRST cohomology  has very different properties when it is computed in the  space of  local or non--local functionals   \cite{Barnich:2003tr}. This is well illustrated by the proof of the non-physical nature of some global symmetries observed in the gauge-fixed action of local Yang--Mills theory in the Curci-Ferarri gauge  \cite{Esole:2003ke} and the difficulties  to define a gauge-independent way to renormalize these non-local operators that take a local expression in particular gauge \cite{Esole:2004jd}.}. 
 The  topology of the configuration space is very important to understand low-energy physics of Yang--Mills theories. Therefore  gauge-invariance should be kept manifest as far as possible  since  gauge-fixing conditions can constrain the value of boundary terms and therefore interfere with the topology. A better understanding of the non-local massive  model based on the functional \Amin calls for  a tractable gauge-invariant formulation which takes into account the consequences of  the  non-local nature of the gauge-invariant theory. Providing such a description   is the  aim of this letter.

We show that the  non-local   massive
Yang--Mills action associated with the operator introduced    by
Gubarev, Stodolsky, and Zakharov is classically equivalent to  a  Yang--Mills principal
gauged sigma model also known as the  Stueckelberg model
\cite{Ruegg:2003ps,Henneaux:1998hq}.   
In this  reformulation, the operator \Amin  corresponds to   the so-called {\em topological term} of the sigma model. 
The non-local action  contains implicitly new degrees of freedom. 
 Locality emerges once   these  degrees of
 freedom  are   explicitly implemented in the action. They corresponds to the  group elements in the sigma model action.  
Conversely, the non-local action is retrieved  from the sigma model
 once the   group elements are  integrated out. 
 A similar scheme is familiar in the Schwinger model  in    $1+1$-dimensions :  
 a  non-local mass term appears once 
the fermionic field is integrated 
 out. 
The sigma model formulation   provides a local theory that  encodes many effects coming from the  non-locality of the original model in a gauge-invariant way.
The  local and non-local formulation are equivalent only at  the classical level. 
 We shall see that the local theory fulfilled the requirement of tree unitarity and is `renormalizable in the modern sense'. These two properties are lost in the non-local model. The sigma model formulation is valid under a certain ultra-violet cut-off which is determined so that the loop corrections are irrelevant in comparison to the tree level amplitudes.  
 The classical reformulation  helps to understand many features of the non-local theories in a gauge-invariant way. Moreover, it provides a safe framework in which calculation are tractable. We shall see that the non-local massive action although quite odd at first look, happens to be  naturally related  to two  more conventional approaches of low-energy QCD. 
The sigma model is relevant in the chiral effective theory of low-energy QCD \cite{Harada:2003jx} which describes gluons and the lightest quarks at low-energy QCD. 
 Interestingly enough, the Stueckelberg action  admits  a  type of solutions which  correspond in the usual Yang-Mills theory  
to configurations called {\em center vortices} 
\cite{Cornwall:1979hz}. The latter  are familiar  in different  (de)-confinement and
chiral symmetry breaking scenarios \cite{deForcrand:1999ms,Engelhardt:2003wm,Reinhardt:2001kf,Cornwall:1999xw}.
 We shall see that  non-locality also constrains the  topology.

To illustrate our method, we  shall first  treat  the Abelian theory. We then   generalize to the   non-Abelian case. We consider  the simple form of the action in the Landau gauge and  we  obtain a  gauge-invariant formulation by  computing its  {\em
  gauge-invariant extension}  \cite{Esole:2004jd}. That is, we
  obtain  a  gauge-invariant action that  reduces to $\int 
  F^2+m^2 A^2$ in the Landau gauge. Gauge-invariant extension is a
 general   technique  that has been
  recently applied to  non-local  Yang--Mills theories \cite{Esole:2004jd} and    effective field theories for  massive gravitons  \cite{Arkani-Hamed:2002sp}.

\section{Sigma model reformulation in the Abelian case}
The action of QED:   
$S[A_\m]=-\frac{1}{2e^2}\mes F^2$ (where  $F^2=\frac{1}{2}
F^{\m\n}F_{\m\n}$ and $F_{\m\n}=\pa_\m
A_\n-\pa_\n A_\m$) is invariant under  an irreducible Abelian gauge symmetry $ \d_\e A_\m = \pa_\m \e$ which is not compatible with the trivial mass term $\int A^2$.  The gauge-invariant extension of the trivial mass term $\int A^2$ from  the Landau gauge is the non-local functional 
$\int \left( A^2+\pA\, \frac{\pA}{\square}\right)$ \cite{Esole:2004jd} 
. This expression  reduces to $\int A^2$  in the Landau gauge ($\pa\cdot
A=0 $)  and corresponds exactly  to the  non-local  mass term of the Schwinger model in $1+1$-dimensions.
 The non-local massive action is    
 \begin{equation}\label{nonLocalAction}
S[A_\m]=-\frac{1}{2e^2}\int F^2-m^2\left( A^2+\pA\, \frac{\pA}{\square}\right).
\end{equation}
This type of non-locality due to the inverse of differential operators
(like  $\square^{-1}$)  is called  ``derived non-locality''  because
it generically comes from integrating out some fields in a local
theory \cite{Eliezer}. Therefore, one  recovers locality by
introducing a new field $\varphi$ in such a way that the new action
$S[A_\m,\varphi]$  reduces to the non-local one 
after $\varphi$ has been replaced by solving its equation of motion. The local-casting gives  here a  Stueckelberg model \cite{Ruegg:2003ps}
\begin{eqnarray}\label{StueckAction0}
S[A_\m,\varphi]\!\!& =&- \frac{1}{2e^2}\!\! \int F^2-
m^2(A^2-2 A^\m \pa_\m\varphi
+\pa^\m\varphi\pa_\m\varphi)\nonumber\\
& =& -\frac{1}{2e^2}\int F^2-
 m^2\left(A_\m-\pa_\m\varphi  \right)^2.
 \end{eqnarray}

 The  gauge  transformation of the scalar field $\varphi$ is  a  shift symmetry
 $\d_\e\varphi=\e $. That symmetry  has to be  spontaneously broken in a quantum theory  since it cannot  be  realized in a  unitary way as a symmetry of the vacuum state.
The Stueckelberg action can be written
 explicitly as a principal  gauged sigma-model  with the circle $S^1$ as target space  
\begin{equation}\label{SigmaModel}
S[A_\m,\phi]=-\frac{1}{2e^2}\int\;  F^2 -m^2\, \cD_\m \phi( \cD^\m\phi)^\dagger,
\end{equation}
where  $\cD_\m X =\pa_\m  X-i\,A_\m X$ and $\phi=e^{i
   {\varphi}   }$. The shift symmetry becomes  a shift of phase : $\delta_\e
 \phi=i\e\phi$. 

An elegant way to understand the  connection between the Proca action and the sigma model  is to use the theory of  constrained system.
It is easy to see  that the Proca  action is a gauge fixed version of the  Stuekelberg  action. 
As usual, fixing the gauge  replaces  first-class constraints by second-class ones. Similarly, an Hamiltonian analysis of the Proca action revels that it possesses second-class constraints. The Stueckelberg action is obtained by trading these second-class constraints for  first class ones. This  implies  the addition of a compensating  field. The  gauge transformation of $\phi$  is  generated by the first class constraints.  Therefore  it  is often stated  that ``the Proca action possesses a hidden symmetry''. The  replacement of a second class constrained system by an equivalent first class one is a general method nowadays known  under the name of {\em conversion}. It is used for the quantification of  theories with second class constraints 
 See  \cite{Batalin:2001je} and reference therein.

\section{Generalization to  Yang--Mills theory}
The free Yang-Mills action  is $S=\frac{1}{g^2}\mes \tr F^2$, where
$g$ is the coupling constant. Our conventions are $F_{\m\n}=\pa_\m
A_\n-\pa_\n A_\m+[ A_\m, A_\n]$ with  $A_\m=A_\m{}^a\ T_a$ where
$[T_a,T_b]=f_{ab}{}^c T_c$, $\tr(T_a T_b)=-\frac{1}{2}\d_{ab}$, and $T_a{}^\dagger=-T_a$.

 In the Abelian case,   the Stueckelberg action $S[A_\mu,\varphi]$ can be  obtained  from the Proca action
 $\frac{1}{2e^2}\int F^2-m^2 A^2$  by applying
 the Stuckelberg transformation $A_\mu\rightarrow
 A_\m-\pa_\mu\varphi$. 

The non-Abelian generalization of the Stueckelberg transformation 
is with our convention $A_\m\rightarrow A'_\m= A_\m-U(x)\inv\pa_\m
U(x)$, where the matrix $U(x)$ is defined in term of the Stueckelberg
variables $\varphi^a(x)$ as $U(x)=\exp\left[\varphi^a(x)
  T_a\right]$. By applying the  Stueckelberg transformation to the
Proca action $S =\frac{1}{g^2}\mes\tr\left[ F^2-m^2 A^2\right]$ we get 
\begin{equation}\label{SA}
S[A_\m,U]=\frac{1}{g^2}\mes\tr \left[F^2-m^2 ( A_\m-U^{-1}\pa_\m U)^2\right].
\end{equation} 
This action is invariant under the finite  gauge transformations 
$A_\m\rightarrow V^{-1}(A_\m+\pa_\m)V$,  $U\rightarrow U V,$ where   $V=\exp[{\e^a(x)T_a}]$. All the non-polynomial sectors of the action depend on $U(\varphi)\inv\pa_\m U(\varphi)$.
Using the   covariant derivative    $\cD_\m U=\pa_\m U-U A_\m$, the previous action can be rewritten in the more familiar  sigma model formulation   
\begin{eqnarray}\label{SM}
S[A,U] =\frac{1}{g^2}\int\!  \tr\left[F^2+m^2\cD_\m U(\cD^{\m} U)^{\dagger}\right]. \end{eqnarray}

The  non-local operator \Amin  corresponds here to    the so-called {\em topological term} of the sigma model 
\begin{equation}\label{masst}
\parallel\!\!\cD
  U\!\!\parallel^2\equiv -\int \tr \left[\cD U (\cD U)^\dagger \right]=-\int \tr(A_\m-U\inv\pa_\m U)^2.
\end{equation}
The equations of motion of the sigma model  are 
\begin{eqnarray}
D^\n F_{\n\m}+m^2\, [A_\m-U\inv\pa_\m U] & =& 0,\label{eom1} \\
m^2\, D^{\m}[A_\m-U\inv\pa_\m U]  &=& 0 \label{eom},
\end{eqnarray}
where $D_{\m}X=\pa_\mu X-[A_\m,X]$ is the (adjoint)-covariant derivative.
These two equations are not independent as the second  can be derived
from the  first one  by acting with the adjoint-covariant
derivative.

If the matrix  $U$ is expanded in terms of the field $\varphi$, one
can easily solve the equation of motion of $\varphi$ \eqref{eom}  in term of the Yang-Mills field $A_\m$ 
 order by order in  
$\Phi\equiv\{A_\m,\varphi\}$ using  
\begin{eqnarray}\label{kformula}
U(\varphi)\inv\pa_\m U(\varphi)=
\sum^\infty_{n=0}\!\frac{(-)^n}{(n+1)!}\left[ad\, \varphi\right]^n
\pa_\m\varphi, \end{eqnarray} 
where $ad X(Y)=[X,Y]$. 
 We obtain a  non-local expression for $\varphi$  which vanishes in
 the Landau gauge and starts like \cite{Delbourgo:1986wz}
\begin{equation}
\varphi= 
\frac{\pa\cdot A}{\square}
- \, \!\frac{1}{\square}
\left[\frac{1}{2}
[ \pa\cdot A,\frac{\pa\cdot A}{\square}]
+\![A^\m, \pa_\m\frac{\pa\cdot A}{\square}]
\right]\!
+{O}(A^3)
\end{equation}

Using the equations \eqref{masst} and  \eqref{kformula} we get the non-local mass term as a funtional of $A_\m$
\begin{equation}\label{varphiA}
\int \tr \left[ A^2+\pa\cdot A \frac{\pa \cdot A}{\square}
+A^\m \left[\pa_\m\frac{\pa \cdot A}{\square},\frac{\pa\cdot A}{\square}\right  ]\right]
+{O}(A^4).
\end{equation}

The Abelian result is recovered when the structure constants vanish.

\section{Discussion and Conclusion}
We shall present two point of view which comes naturally from the reformulation of the effective action as a sigma model. The sigma model fits naturally in the  description of gluons and the lightest quarks  for low energy QCD in the spirit of the chiral effective theory. It can also be viewed as an effective action in  the center vortices description of the features of low-energy QCD. 
We shall also discuss the shortcoming of our methods. This is mainly related to the existence of Gribov copies. We shall see that non-locality constrains the topology. 
\subsection{The Perturbative Chiral Effective Action Approach.}
The sigma model approach   shows that the massive Yang--Mills action
associated with the non-local operator \Amin is  essentially the same
as the model introduced as an alternative to the Higgs model  by Delbourgo and Thompson \cite{Delbourgo:1986wz}. 
However, it has  been found not to be satisfactory as a fundamental theory  because of problems with unitarity \cite{Kubo:1986ps}.   
This explains why we considered the non-local  massive action as an effective  action for low-energy QCD.
 This point of view makes sense as unitarity problem can be disregarded once an appropriate cut-off is considered. 
 In the spirit of effective action  \cite{Gomis:1995jp} it is interested  to look at all the possible local terms that are consistent with the symmetries of the action. Indeed, they might be used as corrections to the action. For the  principal gauged sigma model this question  has been carefully analyzed by  Henneaux and Wilch \cite{Henneaux:1998hq}: in spite of the usual  curvature terms  coming from pure Yang-Mills theory there are also  `winding number terms'  that depend on the group elements and cannot be eliminated when the topology is non-trivial.

The Stueckelberg  fields $\varphi^a$,  play the role of  {unphysical  Goldstone-boson
  fields} since they  decouple  in the so-called {\em
  unitary gauge} ($\varphi
=0 \iff U= I$). The unitary gauge  is equivalent to   the Landau gauge thanks to  equation \eqref{eom}. 			
If we are not in the  Landau gauge (which corresponds here to the unitary gauge $\varphi\neq 0$), equation \eqref{kformula}  shows clearly that  they are an infinite number of non-renormalizable vertices involving $(ad\, \varphi)^n\pa_\m \varphi$. However,  if we were working only in the Landau gauge, we would have the impression that the action is power counting renormalizable. The gauge invariant analysis shows  that renormalizability in the Landau gauge  is a gauge artifact. However,  although the sigma model is not  power counting renormalizable  in 
four spacetime dimensions it is renormalizable in the modern sense of Gomis and Weinberg \cite{Gomis:1995jp} as shown in \cite{Henneaux:1998hq}. That is, all the infinities can be regulated by   counterterms respecting the symmetries of  the original action. But we might need an infinite number of them. Renormalizability in the modern sense is specially well appropriate in the analysis of effective actions \cite{Gomis:1995jp}. 

It  is interesting to  write  equation \eqref{eom} at first order in  the fields $\Phi\in\{A_\m,\varphi \}$
\begin{equation}
\label{long}\pa\cdot A=\square\varphi +{\mathrm
  O}(\Phi^2) \Longrightarrow 
\pa_\m\varphi^a=\pa_\m\frac{\pa\cdot A^{a}}{\square}+{\mathrm
    O}(\Phi^2).
\end{equation} 
 It follows that  the  asymptotic field associated to the
 longitudinal mode of the $A_\m$ is just equivalent to $\pa_\m
 \varphi$. Therefore,   the latter can validly replace the former
 when computing $S$-matrix elements   at  tree level on the mass-shell  in the 
  limit in which the energies are all much greater than the vector
  boson mass. This is a direct application of the Equivalence Theorem  \cite{He:1993qa}.

The principal gauged sigma  model is the starting block of the chiral perturbative theory of QCD. The latter provides  an effective action of gluons and lightest hadronics fields. It  describes the  chiral symmetry properties of  QCD in the light of effective field theory. The central phenomena is here the spontaneous chiral symmetry breaking which admits  the quark condensate as an order parameter.  
The effective theory breaks down at the scale $\Lambda$ such that the loop correction becomes relevant. The naive dimensional analysis (NDA) gives   $\Lambda\sim 4\pi f$ where $ f=\frac{ m}{g}$. In the spirit of large $N$- limit calculation, the cut-off can be improved to  $\Lambda_N\sim\frac{ 4\pi f}{\sqrt{N}}$   by taking into account the number of colors ($N$) in  coefficients of  loop corrections   \cite{Harada:2003jx}.
The  cut-off is chosen so that  loop corrections are negligible compare to  tree-level amplitudes. Therefore, it  also protects the model from unitarity violation since the latter are only seen at loop-levels. 
 It is amusing that from a non-local action which involves only Yang--Mills field we end up with an action that aims to describe hadronic matter as well. 

\subsection{The center vortices point of view.}
The equations  of motion  of the Stuckelberg action admit non-trivial
topological  solutions as was emphasized particularly by Cornwall
\cite{Cornwall:1979hz}. 
  Their  non trivial topological features
can already be seen by noticing that the second equation of motion \eqref{eom} 
 is just  the   Landau  background gauge condition,  which is well known not
to have a unique solution  when physical boundary conditions are
considered. These Gribov ambiguities are a  manifestation of the
non-trivial topology of the configuration space of non-Abelian gauge
theories and  also  occur in QED when the gauge group is compact and spacetime has a finite volume.  
 One can see that static solitonic excitation are possible  by using a familiar  rescaling  argument in the static action. After a rescaling   $x\rightarrow \lambda x$, the dynamical term $\int dx^3 F^2$  scales like $\lambda$ whereas the topological term  $\int dx^3  \parallel \cD_\m \parallel^2$ scales like $1/\lambda$. Therefore the sum of the two terms always has a minimum with $\lambda\sim m$ \cite{Cornwall:2001ni}. 

 An important  type of  solutions of the equations of motion of the
 Stueckelberg action are the so-called   {\em center vortices} \cite{Cornwall:1979hz,Cornwall:2001ni}.  
They  are   localized gauge field configurations which carry flux
 concentrated on a  closed hypersurfaces  of co-dimension two ( sheet
 in 4 D , loops in 3D, point in 2D) and  many of their  properties
 are encoded in the topological features of the hypersurfaces like the
 intersection number and  the linking number\cite{Reinhardt:2001kf,Cornwall:1999xw,Cornwall:2001ni}. Center vortices  have the nice property of being described and analysed in a gauge-invariant way \cite{Cornwall:2001ni,Reinhardt:2001kf}. 
Center vortices have  attracted a lot of attention following famous 
lattice simulations which  indicate that  their  removal  from the grand ensemble destroys confinement and  chiral
symmetry breaking and eliminates all the topological non-trivial
field configurations \cite{deForcrand:1999ms,Engelhardt:2003wm}. However,  the analytic confirmation or refutation of these scenarios  is still missing in the continuous formulation of QCD.

 \Amin  was first introduced to study topological structures in
 gauge theories. This  was illustrated by a lattice simulation which
  indicates that the expectation value of \Amin is sensitive to the
 phase transition of compact QED \cite{Gubarev:2000eu}. It was also argued on the lattice that the expectation value of the square potential is associated to instantons in the Landau gauge  \cite{Boucaud:2002nc}.

The expectation value of \Amin can also  be  formulated in terms   of a principal gauged sigma model.
Indeed, the  expectation value of an operator $\cO$ in a model 
 defined by an action $S(\Phi)$ is defined by $\langle\cO\rangle =
  \int d\Phi\, \cO(\Phi)\,\exp\left[ S(\Phi)\right]$. It is useful to  rewrite
 it in the following way
\begin{eqnarray}
 \langle\cO\rangle =
 \frac{\pa}{\pa \lambda} \left.{ \int d\Phi\exp\left[
    S(\Phi)+\lambda \cO(\Phi)\right]\,}\right|_{\lambda=0}
\end{eqnarray} 
where the functional $
 S(\Phi)+\lambda \cO(\Phi)$ is called  the  {\em  effective
  action for the operator $\cO(\Phi)$}. This is  a very practical concept. For example,   the correlation functions
$\langle\phi(y_1)\cdots\phi(y_r)\cO(z)\rangle$ are just the first
order $\lambda$-coefficients in the usual Schwinger functions for the
action  $ S(\Phi)+\lambda \cO(\Phi)$. 
The effective action  for the operator \Amin is  the non-local massive
 Yang-Mills  action $F^2+\lambda \Amin$, which can be reformulated  as a  
 principal gauged sigma model  of equations \eqref{SA} and \eqref{SM}. 
In view of the relation between the sigma model and center vortices \cite{Cornwall:1979hz}, it is natural to ask if \Amin is sensitive to center vortices configurations. It would be possible to study this possiblity with a lattice simulation.

\subsection{Local vs non-local formulation.}
The non-local  massive Yang--Mills theory having as a mass term the minimum of the square potential along the gauge potential  is  equivalent to a gauged-principal sigma model in the following sense.
The non-local theory is obtained from the sigma model by integration out the group elements present in the sigma model action. In the other direction, the sigma model emerges once the degrees of freedom implicitly generated by  non-locality are explicitly included in the action.

Although  the local and non-local formulation are equivalent in the sense explain above, they do  have important differences. For example,  the Stueckelberg model  satisfies  the tree unitarity conditions \cite{Cornwall:1974km} 
\footnote{Tree unitary : The $n$-particle $S$-matrix elements do not grow more rapidly than $E^{4-N}$ in the limit $E\rightarrow \infty$. See \cite{Cornwall:1974km} for more details.}
 and is renormalizable in the modern sense of Gomis and Weinberg
\cite{Gomis:1995jp}
\footnote{That is, all the infinities can be eliminate by using local counter-terms that respects the symmetries of the original action, however, one might needs an infinite number of them.}
 as shown in  \cite{Henneaux:1998hq}. This two properties disappear in the non-local theories  \cite{Kubo:1986ps,Esole:2004jd}. This can be understood from the fact that the transformation from one model to the other is highly non-local whereas tree  unitarity and renormalizability in the modern sense are mostly valid for (perturbatively) local  theories. Quantum correction can be seen as part of the measure of the path-integral. The latter is known to be invariant under field redefinitions. However, when non-locality is introduced, the physics can be altered. This is well illustrated by the re-investigation  of the Equivalence Theorem by Tuytin \cite{Tyutin:2000ht}. It is shown there that  two field theories related by a field redefinition  can have different physical content even when the field redefinitions are local or perturbatively local.

\subsection{The gauged-principal sigma model as a collective description  of the local minima of the  square-potential.}
The sigma model  that we present in this letter is the gauge-invariant extension of the trivial massive YM theory from the Landau gauge. 
Therefore its equivalence to the  gauge-invariant definition of the \Amin-massive YM action (which requires the computation of the absolute minimum along the gauge orbit) relies  on the ability of the  Landau gauge to detect the absolute minimum. 
However, the  Landau gauge condition does not single out a unique value of the potential $A_\m$  as it pocesses Gribov copies.
 That is, there exists different gauge fields related by a gauge transformation but satisfying all the Landau gauge condition.
 The Landau condition   characterizes local minima of the square potential along the gauge orbit. The  absolute  minimum among all possible Gribov copies is never selected.  Since the sigma model is obtained from the Landau gauge description,  it   inherits the Landau gauge ambiguities :  
  the sigma model  encodes only information associated with local minima of the  square potential and does not identify the global minimum.  
Still it is the gauge-invariant formulation of the `Landau gauge dynamical mass generation' as  discuss in the literature. Indeed, the  latter is only defines by the Landau gauge-fixed action  and therefore suffer as well of the  Gribov ambiguities.  
 The difficulties  of identifying the absolute minima of the square potential is a common feature of all  the methods used  to evaluate \Amin.
The problem of Gribov copies in the definition of \Amin is carefully  analyzed in \cite{Stodolsky:2002st} to which we refer for more details.
 The ambiguities in the definition of \Amin  are not only a  technical problem. They are related to miscellaneous properties of the configuration space of Yang--Mills theory. They depend  on  particular features of the gauge potential that we start with. In particular, they are unavoidable when we deal with  reducible gauge potential \cite{Fuchs:1995mb}.

We would like as well to point that topological consideration are crucial to make sense of the non-local  mass term \eqref{nonLocalAction}, \eqref{varphiA}. Taking the inverse of  a differential operator   makes sense only if the latter  has no  non-trivial zero
modes. The possible   boundary
conditions are therefore constrained. For consistency the  set
of gauge transformations has to be limited to those that  preserve
 these boundary conditions.  In the Hamiltonian analysis  of the
 operator \Amin,   restrictions are imposed   in the Landau
  gauge to avoid  Gribov 
  copies  \cite{Stodolsky:2002st}.

 We do not think that a  model that determines the exact absolute 
minimum can be  defined in a tractable way. When the connection field $A_\m$ is reducible, the spectrum of absolute minima is degenerated \cite{Fuchs:1995mb}.   
 In this respect, the sigma model reformulation is a fair   alternative. 
 It   is local, manifestly gauge-invariant, renormalizable in the modern sense, and  fulfills the tree unitarity condition. It provides a gauge-invariant description in which calculations are
 possible while the  effects of non-locality are taken into account  through the  additional fields.
 The gauged principal sigma model is also  interesting  in its own without any link with the \Amin operator.  
 At low energy the gauged principal  sigma model  can be used as an effective action for gluons and lightest hadrons in the spirit of perturbative  chiral effective theory.  In the spirit of the  center vortices pictures of low energy QCD
 an effective action relevant to understand confinement and chiral symmetry breaking.
 In view of these features of the gauged principal sigma model, it would be interesting to study the sensitivity of the  expectation value of \Amin to center vortices configurations.
 
\subsection*{ Acknowledgements.}  
 It is a pleasure to thank A. {Ach\'ucarro}, P. van Baal, F. Bruckmann,  M. Bucher,  M. Henneaux, M.O. de Kok, B. Leurs, M. Pickles, and H. Verschelde for
useful discussions. We are specially greatful to F. Freire for a  careful reading of the manuscript and  discussions.

We would like to thank as well the Staff and the students of the 
African 
Institute for Mathematics, where this work has been done for 
their hospitality and the  Ford Fellowship for financial support.

\end{document}